\documentclass{aastex63}
\received{Sep. 10, 2021}
\revised{Nov. 07, 2021}
\accepted{Nov. 8, 2021}
\shorttitle{LLAGN with SFPR}
\shortauthors{Jiang et al.}

\begin{document}

\title{Millimeter-VLBI observations of low-luminosity active galactic nuclei with source-frequency phase-referencing}

\correspondingauthor{Wu Jiang.}
\email{jiangwu@shao.ac.cn}

\author[0000-0001-7369-3539]{Wu Jiang}
\affiliation{Shanghai Astronomical Observatory, 
 Chinese Academy of Sciences, 
 Shanghai 200030, China}
\affiliation{Key Laboratory of Radio Astronomy,  
 Chinese Academy of Sciences, Nanjing 210008, China}
 
\author[0000-0003-3540-8746]{Zhiqiang Shen}
\affiliation{Shanghai Astronomical Observatory, 
 Chinese Academy of Sciences, 
 Shanghai 200030, China}
\affiliation{Key Laboratory of Radio Astronomy,  
 Chinese Academy of Sciences, Nanjing 210008, China}
 
\author[0000-0003-3708-9611]{Ivan Mart\'i-Vidal}
\affiliation{Department of Astronomy and Astrophysics, 
University of Valencia,
E-46100, Burjassot, Spain }

\author{Xuezheng Wang}
\affiliation{Shanghai Astronomical Observatory, 
 Chinese Academy of Sciences, 
 Shanghai 200030, China}
\affiliation{ShanghaiTech University,
393 Middle Huaxia Road, Pudong, Shanghai, 201210, 
China}
 
\author{Dongrong Jiang}
\affiliation{Shanghai Astronomical Observatory, 
 Chinese Academy of Sciences, 
 Shanghai 200030, China}
\affiliation{Key Laboratory of Radio Astronomy, 
 Chinese Academy of Sciences, Nanjing 210008, China}

\author[0000-0002-7776-3159]{Noriyuki Kawaguchi}
\affiliation{National Astronomical Observatory of Japan, 
 2-21-1 Osawa, Mitaka, Tokyo 181-8588, Japan}



\begin{abstract}
We report millimeter-VLBI results of low-luminosity active galactic nuclei (M\,84 and M\,87) up to 88\,GHz with source-frequency phase-referencing observations. We detected the weak VLBI core and obtained the first image of M\,84 at 88\,GHz. The derived brightness temperature of M\,84 core was about 7.2$\times10^9$\,K, which could serve as a lower limit as the core down to 30 Schwarzschild radii was still un-resolved in our 88\,GHz observations. We successfully determined the core-shifts of M\,87 at 22-44\,GHz and 44-88\,GHz through source-frequency phase-referencing technique. The jet apex of M\,87 could be deduced at $\sim$46\,$\mu$as upstream of the 43\,GHz core from core-shift measurements. The estimated magnetic field strength of the 88\,GHz core of M\,87 is $4.8\pm2.4$\,G, which is at the same magnitude of 1-30\,G near the event horizon probed by the Event Horizon Telescope.    

\end{abstract}



\section{Introduction} \label{sec:intro}

The low-luminosity active galactic nuclei (LLAGNs) classified by their low bolometric
luminosities and sub-Eddington accretion rate commonly exist in nearby galaxies \citep{2002A&A...392..53N, 2008ARA&A..46..475H}. Unlike their bright cousins, the broad-band spectral energy distributions prefer the model of an inner advection-dominated accretion flow and an outer truncated thin disc \citep{2011ApJ...726...87Y, 2014MNRAS.438.2804N}. Meanwhile compact flat-spectrum radio cores were detected in LLAGNs \citep{2002A&A...392..53N} and suggested to be scaled-down versions of AGN jets \citep{1999A&A...342...49F}. Hosting a large mass of central supermassive black hole and its proximity make LLAGN approachable to the launching and accelaration region of  inner jet, even to its event horizon at millimeter or sub-millimeter wavelength.  \\
According to the inhomogeneous model of relativistic jet, the position of the VLBI core is frequency-dependent. This frequency-dependent shift in the location of the core (core-shift) can be used to estimate the magnetic field strength and electron number density of the jet \citep{1998A&A...330..79L,2008A&A...483..759K}. However, only a few active galactic nuclei (AGNs) have reliable core-shift measurements at millimeter wavelengths \citep{2009MNRAS...400..26O}. The core-shift measurements are mostly obtained at low frequencies due to limited sensitivity at high frequencies  \citep{2012A&A...545..133P}. Although millimeter-VLBI can approach the inner region of jet as the plasma turns optically thin at high frequency, the core-shift is difficult to obtain due to the rapid phase flutuations of atmosphere and thus limited coherent intergration time for the conventional VLBI phase-referencing observations. Fortunately, a newly proposed VLBI phase-referencing technique called source-frequency phase-referencing (SFPR) can be used to measure the core-shift effect \citep{2011AJ....141..114R}, essentially being of great advantadge at  millimeter wavelengths. We successfully obtained the first VLBI image of the LLAGN M\,81* at 88\,GHz and measured the shift between 7\,mm and 3\,mm wavelengths in the compact jet \citep{2018APJL...853..14J}. \\
In this paper, we will present the applications of SFPR to two LLAGNs, M\,84 and M\,87. The observation summary and data reduction are presented in section 2. The results are in section 3, followed by the conclusion in section 4.

\section{Observations and data reduction} \label{sec:Data}

\subsection{Observations \label{obs}}
The observations of M\,84 and M\,87 were carried out with the VLBA in SFPR mode on June 22, 2019 with 22 and 44\,GHz frequency pair. The observations with 44 and 88\,GHz frequency pair were firstly performed on June 18, 2019 but most stations failed at 88\,GHz. Therefore it was reobserved on March 31, 2021 for satisfied weather conditions at most stations. The on-source time of each source per scan was $\sim$30 seconds at a frequency. A fast frequency-switching cycle of two frequencies on a source was taken as a loop. M\,87 ($\sim1^\circ.5$ apart from M\,84 in the sky) was interleaved  every 5 loops of M\,84 for the 22/44 \,GHz epoch and 9 loops for the 44/88\,GHz epoch. Several loops of blazar 1219+044 were observed at the beginning and the end of observations but not included here, due to rapid phase flutuations of low antenna elevation angles. The 2019 epoch was recorded at 2048\,Mbps by splitting the 512\,MHz total bandwidth into 16 intermediate frequency (IF) bands while the 2021 epoch was at 4096\,Mbps by splitting 1024\,MHz bandwidth into 8 IFs. 
\subsection{Data reduction \label{reduction}}
The data calibration and reduction followed the procedures in \citet{2011AJ....141..114R} and \citet{2018APJL...853..14J}. Firstly we performed standard amplitude and phase calibrations in AIPS for both M\,84 and M\,87 at the reference frequencies (22\,GHz of 2019 epoch and 44\,GHz of 2021 epoch), respectively. Their images were obtained by further clean and self-calibration in Difmap. Secondly the corresponding phase solutions of the AIPS task FRING after taking into account clean models of M\,84 and M\,87 were multiplied by a factor of two, while the delay and delay rate solutions were unchanged. These revised solutions were then applied to the target frequencies (44\,GHz in 2019 epoch and 88\,GHz in 2021 epoch, respectively), which is called the frequency phase transfer (FPT) calibration. The phase flutuations in proportion to the observing frequency such as the unmodeled tropspheric and geometric errors were eliminated in this step. Thirdly the SFPR-ed images of M\,84 at target frequencies were obtained by further phase referencing to the FRING solutions of the FPT calibrated data of M\,87 at the same target frequencies, which refined the unmodeled dispersive ionospheric and instrumental errors. The corresponding core-shift measurements were derived from the SFPR-ed images using JMFIT in AIPS. Finally high resolution VLBI images of M\,84 were obtained by clean and self-calibration in Difmap \citep{2018APJL...853..14J}. \\
Since the brightness-peak position of the image is usually referred as the core position, the prominent jet of M87 in right ascension (RA) direction would cause the peak position to be slightly offset from the ``true core" position towards the downstream side, due to the blending of near-core jet emission within the finite beam size. To evaluate this effect, we used the method in \citet{2014ApJ...788..165H}, the M\,87 structure was convolved by different beam size with diameters ranging from the minor axis of the nominal synthesized beam (shown in Table 1), whose direction was almost in RA direction in our observations, to about four times larger. Then, we plotted systematic changes of the brightness-peak position as a function of beam size. In the case of 22\,GHz, the image were restored with beam sizes ranging from 0.3 to 1.2\,mas in incremental steps of 0.1\,mas. We found a progressive position shift of the brightness peak toward the downstream to be 10\,$\mu$as per 0.1 mas in RA direction. In the case of 44\,GHz, the image was restored with beam sizes ranging from 0.2 to 0.8\,mas in incremental steps of 0.1\,mas. The position shift of the brightness peak toward the downstream was 6.8\,$\mu$as per 0.1\,mas in RA direction. In the case of 88\,GHz, the image was restored with beam sizes ranging from 0.15 to 0.6\,mas in incremental steps of 0.05\,mas. The position shift of the brightness peak toward the downstream was 2.5\,$\mu$as per 0.1\,mas in RA direction. That means the ``true core'' position would be shifted to upstream, with respect to the brightness-peak position when convolved with a nominal beam. At 22\,GHz, the upstream shift in RA direction would be about 10$\times$0.39\,mas/0.1\,mas$=39$\,$\mu$as. At 44\,GHz, it would be about 6.8 $\times$ 0.22\,mas/0.1\,mas $= $15\,$\mu$as for the 2019 epoch, and about 6.8$\times$0.24\,mas/0.1\,mas $= 16$\,$\mu$as for the 2021 epoch. At 88\,GHz, it was about 2.5$\times$0.16\,mas/0.1\,mas ${=} $4\,$\mu$as. Where, 0.39\,mas, 0.22\,mas, 0.24\,mas and 0.16\,mas were the nominal beam sizes in the RA direction (see Table 1). Consequently, the measured core shift in RA direction from the peak-positions would be about 39$-15=24$\,$\mu$as larger than the ``true core shift'' in magnitude at 22-44\,GHz and, 16$-4=$12\,$\mu$as larger at 44-88\,GHz. 
\subsection{Error analysis \label{error}}
The error budgets of SFPR mainly include the dynamic tropospheric error, the core identification error of M\,87 and the statistical error from images. These errors are independently to each other and the total errors can be calculated as the root-sum-square of them. We adopt 11\,$\mu$as for the dynamic tropospheric error among frequencies under relatively stable weather conditions, assuming 0.01\,m uncanceled error by the water vapor fluctuation as that in \citet{2011Natur...477..185H}. The absolute tropospheric position error for a single frequency can be significantly larger than this value, while most of the error can be canceled out by the FPT calibration in SFPR observations. We also performed the error analysis for the core identification error of M\,87 as in \citet{2011Natur...477..185H}, using the core position differences between two methods. One defined the centroid of the elliptical Gaussian fitting M\,87 core region as the core position, the other is the brightness-peak position of the images convolved with a circular Gaussian beam of about a half of synthesized beam in the core-jet direction. The uncertainties of core identification of M\,87 in RA direction are 7\,$\mu$as, 6\,$\mu$as at 22\,GHz and 44\,GHz for the 2019 epoch, respectively. It was 5\,$\mu$as, 3\,$\mu$as at 44\,GHz and 88\,GHz for the 2021 epoch, respectively. Since the SFPR-ed images coupled the errors from both frequencies, the statistical error at each frequency was 1/$\sqrt{2}$ of the beam size divided by the signal-to-noise ratios of SFPR-ed image. It was 12\,$\mu$as at 22\,GHz and 44\,GHz for the 2019 epoch, 14\,$\mu$as at 44\,GHz and 88\,GHz for the 2021 epoch. The intrinsic structural uncertainty of M84 would also affect the core-shift of M\,87. In our observations, the position angles of fitting to the core region of M\,84 at 22 and 44\,GHz with an elliptical Gaussian were within 5 degrees in the north. Therefore the uncertainty of core-shift effect in M\,84 in RA direction was 5\,$\mu$as, 3\,$\mu$as and 2\,$\mu$as at 22, 44 and 88\,GHz, respectively. Other minor errors such as the ionospheric residuals and geometric errors at each frequency were taken the empirical values \citep{2011Natur...477..185H}. The total errors by root-sum-square all the above uncertainties give out 19\,$\mu$as and 18\,$\mu$as at 22 and 44\,GHz for the 2019 epoch, 19\,$\mu$as and 18\,$\mu$as at 44 and 88\,GHz for the 2021 epoch, respectively.

\section{Results and discussions} \label{sec:analysis}
\subsection{VLBI core of M\,84 at 3\,mm \label{compact core}}
The nearby elliptical galaxy M\,84 is located in the center of Virgo Cluster at a distance
of 18.5 Mpc ($z=0.00339$) and has a central supermassive black hole weighing $\sim8.5\times10^8 M_\odot$. The combination of its proximity and a large black hole mass yields a privileged linear resolution conversion factor down to 1\,micro-arcsecond ($\mu$as) $\sim$ 1 Schwarzschild radii ($R_S$), allowing us to investigate its close vicinity of the supermassive black hole with VLBI. Two side jets are seen at a large viewing angle of $\sim$74$^\circ$ \citep{2018ApJ...860....9M}. The image of M\,84 at 88\,GHz (Figure 1) was obtained by performing further clean and self-calibration in phase only to the SFPR-ed visibility data, which were scan averaged. The MODELFIT task in Difmap was used to fit the calibrated visibility with circular Gaussian components. The VLBI core of M\,84 could be fitted by a circular Gaussian with a flux density of 38.0$\pm5.7$\,mJy and a diameter of 29$\pm11$\,$\mu$as well. The apparent brightness temperature of the core \citep{2018A&A...616A.188K}, $T_{b, app}$, can be calculated by   
\begin{equation}
T_{b, app} =1.22\times10^{12} \frac{S_{core}(1+z)}{\nu^2\theta_{core}^2} \,K,
\end{equation}
where $S_{core}$ is the core flux density in Jy, $\nu$ is the observing frequency in GHz, $\theta_{core}$ is the equivalent size in milli-arcsecond (mas) and $z$ is the redshift. The $T_{b,app}$ of M\,84 core at 88\,GHz is $\sim7.2\times10^9$\,K. Similar to other LLAGNs \citep{2018A&A...616A.188K}, the $T_{b,app}$ is generally quite low. The derived brightness temperature could serve as a lower limit as the core size of 29\,$\mu$as (30 $R_S$) was only one-tenth of the beam size and still un-resolved in our 88\,GHz observations.

\subsection{Core-shift of M\,87 \label{core-shift}}
M87 is the most prominent elliptical galaxy within the Virgo Cluster, located at a distance of 16.8$\pm$0.8\,Mpc away. Its central supermassive black hole ($\sim6.5\times10^9 M_\odot$) and jet have been well-studied in almost every wave band from radio to $\gamma$-rays \citep{2019ApJ...875L...1E,2021ApJ...911L..11E}. The core-shift of M\,87 up to 43\,GHz has been measured through conventional phase-referencing to M\,84  \citep{2011Natur...477..185H, 2013APJ...775..70H}. The core-shift in RA direction $r_{RA}(\nu)$ followed $\nu^{-0.94}$ and indicated that the black hole was located at $\sim$41\,$\mu$as eastwards of the 43\,GHz core \citep{2011Natur...477..185H}. The core-shifts at 22-44\,GHz and 44-88\,GHz could be obtained from the SFPR-ed images \citep{2011AJ....141..114R, 2018APJL...853..14J} as shown in Figure 2. Since the jet extended structure of M\,84 is toward the north direction, the core-shift of M\,84 in RA direction could be negligible. The core-shift of M\,87 in RA direction could be obtained from the SFPR-ed images by Equation (2) in \citet{2018APJL...853..14J}. As a result, we obtained a position shift in RA direction of $-64\pm8$\,$\mu$as at the 22\,GHz core with regard to the 44\,GHz core, and of $-33\pm11$\,$\mu$as at 44-88\,GHz. The error in 1$\sigma$ was given by JMFIT in AIPS. After taking into account of the blending effect of near-core jet emission in M\,87 (see Section 2.2), the ``true core shift'' would be about $-(64-24)$=$-40$\,$\mu$as at 22-44\,GHz and $-(33-12)$=$-21$\,$\mu$as at 44-88\,GHz. Incorporating the uncertainties given by error analysis (see Section 2.3), the core positions relative to that of the 44\,GHz core in RA was $-40\pm19$\,$\mu$as at 22\,GHz for the 2019 epoch and $21\pm18$\,$\mu$as at 88\,GHz for the 2021 epoch, respectively. \\
Using the same formula $r_{RA}(\nu)=A\nu^{-k}+B$ in \citet{2011Natur...477..185H}, the above-mentioned two core-shifts in RA direction could be solved with solutions $A=-1.45$, $k=0.92$ and $B=0.045$. As presented in the bottom-right corner of Figure 3, assuming a jet position angle of $-69^\circ$ with respect to north in M\,87 \citep{2018A&A...616A.188K}, $B$ value indicates that the jet apex is located at $\sim$48\,$\mu$as upstream of the 43\,GHz core. It was consistent with previous result of $44\pm13$\,$\mu$as in \citet{2011Natur...477..185H}. Since there was no strong flare event to cause the core-shift variations \citep{2019MNRAS.485.1822P} during our observations as well as in \citet{2011Natur...477..185H} session, it would be reasonable to align the 43\,GHz core positions of M\,87 among these sessions. The core-shift in RA at 22-43\,GHz even during the elevated very high energy Gamma-ray state in 2012 was found to be also at a similar level ($\sim$10\,$\mu$as larger) \citep{2014ApJ...788..165H}. Furthermore, aligning the 43\,GHz cores among \citet{2011Natur...477..185H} and the two epochs of this work, we fitted these combined core-shift measurements with weighted least square method. It gave out $A=-1.36\pm0.15$, $k=0.92\pm0.06$ and $B=0.043\pm0.007$ as shown in Table 2. The results are consistent and it implies the jet apex is 43$\pm7$\,$\mu$as in RA (as indicated by black dashed line in Figure 3) or at $\sim$46\,$\mu$as upstream of the 43\,GHz core. \\
Following the Equation (4) in \citet{1998A&A...330..79L}, we could estimate the core-shift measure $\Omega_{r,22-44}=0.13\pm0.06$\,[pc\,GHz] and $\Omega_{r,44-88}=0.12\pm0.10$\,[pc\,GHz], using the resultant power law index $k_r = 1.09$. The magnetic field strength at the 88\,GHz core is estimated to be $4.8\pm2.4$\,G using the mean value of $\Omega_r$, by Equation (B.2) in \citet{2021A&A...650L..18P} with a spectral index of $-0.5$, a Doppler factor of 2, a jet viewing angle of 18$^\circ$ and an intrinsic jet opening angle of $63^\circ.6$ at 88\,GHz \citep{2018A&A...616A.188K}. This is consistent with the estimated 1-30\,G near event horizon by the Event Horizon Telescope at 230\,GHz \citep{2021ApJ...910L..13E}.

\section{Conclusions} \label{sec:conclusion}
We have successfully demonstrated that the SFPR technique could be applied to the mm-VLBI observations of LLAGNs. It helps to overcome the limited coherent intergration time and has great advantages in detecting the weak VLBI core as well as measuring the core-shift at millimeter-wavelengths. The VLBI core of M\,84 at 88\,GHz was detected and the lower limit of its apparent brightness temperature $\sim7.2\times10^9$\,K was obtained. By means of SFPR to M\,84, the core-shift of M\,87 in RA direction was determined at a precision of $\sim$20\,$\mu$as, which further constrained the jet apex at $\sim$46\,$\mu$as upstream of the 43\,GHz core. With the aid of simultaneous multi-frequency receiving system and more stations available \citep{2019JKAS...52...23Z, 2020A&ARv..28....6R}, SFPR will be a very powerful tool to investigate the compactness of the jet base at high frequencies as well as physical parameters such as core structure, brightness temperature and magnetic field of the inner region of jet.

\acknowledgments
The authors thank the anonymous referee for very critical and constructive suggestions. This work was supported in part 
by the National Natural Science Foundation of China (grant Nos. 11803071, 
11590780, 11590784,  and 11933007) and Key Research Program of Frontier Sciences, 
CAS (grant No.QYZDJ-SSW-SLH057). VLBA is operated by the National
Radio Astronomy Observatory, which is a facility of the National
Science Foundation operated under cooperative agreement by
Associated Universities, Inc. 







\begin{thebibliography}{}
\expandafter\ifx\csname natexlab\endcsname\relax\def\natexlab#1{#1}\fi
\providecommand{\url}[1]{\href{#1}{#1}}
\providecommand{\dodoi}[1]{doi:~\href{http://doi.org/#1}{\nolinkurl{#1}}}
\providecommand{\doeprint}[1]{\href{http://ascl.net/#1}{\nolinkurl{http://ascl.net/#1}}}
\providecommand{\doarXiv}[1]{\href{https://arxiv.org/abs/#1}{\nolinkurl{https://arxiv.org/abs/#1}}}
\bibitem[Event Horizon Telescope Collaboration et al.(2019)]{2019ApJ...875L...1E} Event Horizon Telescope Collaboration, Akiyama, K., Alberdi, A., et al.\ 2019, \apjl, 875, L1. doi:10.3847/2041-8213/ab0ec7
\bibitem[EHT MWL Science Working Group et al.(2021)]{2021ApJ...911L..11E} EHT MWL Science Working Group, Algaba, J.~C., Anczarski, J., et al.\ 2021, \apjl, 911, L11. doi:10.3847/2041-8213/abef71
\bibitem[Event Horizon Telescope Collaboration et al.(2021)]{2021ApJ...910L..13E} Event Horizon Telescope Collaboration, Akiyama, K., Algaba, J.~C., et al.\ 2021, \apjl, 910, L13. doi:10.3847/2041-8213/abe4de
\bibitem[Falcke \& Biermann(1999)]{1999A&A...342...49F} Falcke, H. \& Biermann, P.~L.\ 1999, \aap, 342, 49
\bibitem[Hada et al.(2011)]{2011Natur...477..185H} Hada, K., Doi, A., Kino, M., et al. \ 2011, Natur, 477, 185
\bibitem[Hada et al.(2013)]{2013APJ...775..70H} Hada, K., Kino, M., Doi, A., et al. \ 2013, ApJ, 775, 70
\bibitem[Hada et al.(2014)]{2014ApJ...788..165H} Hada, K., Giroletti, M., Kino, M., et al.\ 2014, \apj, 788, 165. doi:10.1088/0004-637X/788/2/165
\bibitem[Ho(2008)]{2008ARA&A..46..475H} Ho, L.~C.\ 2008, \araa, 46, 475. doi:10.1146/annurev.astro.45.051806.110546
\bibitem[Jiang et al.(2018)]{2018APJL...853..14J} Jiang, W., Shen, Z.-Q., Jiang, D.-r., Marti-Vidal, I., \& Kawaguchi, N.\ 2018, ApJL, 853, L14
\bibitem[Kim et al.(2018)]{2018A&A...616A.188K} Kim, J.-Y., Krichbaum, T.~P., Lu, R.-S., et al.\ 2018, \aap, 616, A188. doi:10.1051/0004-6361/201832921
\bibitem[Kovalev et al.(2008)]{2008A&A...483..759K} Kovalev, Y.~Y., Lobanov, A.~P., Pushkarev, A.~B., et al. \ 2008, A\&A, 483, 759
\bibitem[Lobanov(1998)]{1998A&A...330..79L} Lobanov, A.~P., \ 1998, A\&A, 330, 79
\bibitem[Nagar et al.(2002)]{2002A&A...392..53N} Nagar,N.~M., Falcke, H., Wilson, A.~S., et al.\ 2002, \aap, 392, 53
\bibitem[Meyer et al.(2018)]{2018ApJ...860....9M} Meyer, E.~T., Petropoulou, M., Georganopoulos, M., et al.\ 2018, \apj, 860, 9. doi:10.3847/1538-4357/aabf39
\bibitem[Nemmen et al.(2014)]{2014MNRAS.438.2804N} Nemmen, R.~S., Storchi-Bergmann, T., \& Eracleous, M.\ 2014, \mnras, 438, 2804. doi:10.1093/mnras/stt2388
\bibitem[O'Sullivan \& Gabuzda(2009)]{2009MNRAS...400..26O} O'Sullivan, S.~P., \& Gabuzda, D.~C.,\ 2009, MNRAS, 400, 26
\bibitem[Paraschos et al.(2021)]{2021A&A...650L..18P} Paraschos, G.~F., Kim, J.-Y., Krichbaum, T.~P., et al.\ 2021, \aap, 650, L18. doi:10.1051/0004-6361/202140776
\bibitem[Plavin et al.(2019)]{2019MNRAS.485.1822P} Plavin, A.~V., Kovalev, Y.~Y., Pushkarev, A.~B., et al.\ 2019, \mnras, 485, 1822. doi:10.1093/mnras/stz504
\bibitem[Pushkarev et al.(2012)]{2012A&A...545..133P} Pushkarev, A. B., Hovatta, T. ,  Kovalev, Y. Y., et al. \ 2012,  A\&A, 545, 133
\bibitem[Rioja \& Dodson(2011)]{2011AJ....141..114R} Rioja, M., \& Dodson, R., \ 2011, AJ, 141, 114
\bibitem[Rioja \& Dodson(2020)]{2020A&ARv..28....6R} Rioja, M.~J. \& Dodson, R.\ 2020, \aapr, 28, 6. doi:10.1007/s00159-020-00126-z
\bibitem[Yu et al.(2011)]{2011ApJ...726...87Y} Yu, Z., Yuan, F., \& Ho, L.~C.\ 2011, \apj, 726, 87. doi:10.1088/0004-637X/726/2/87
\bibitem[Zhao et al.(2019)]{2019JKAS...52...23Z} Zhao, G.-Y., Jung, T., Sohn, B.~W., et al.\ 2019, Journal of Korean Astronomical Society, 52, 23. doi:10.5303/JKAS.2019.52.1.23
\end{thebibliography}
\bibliographystyle{aasjournal}

\begin{figure}[htb]
\includegraphics[width=6in]{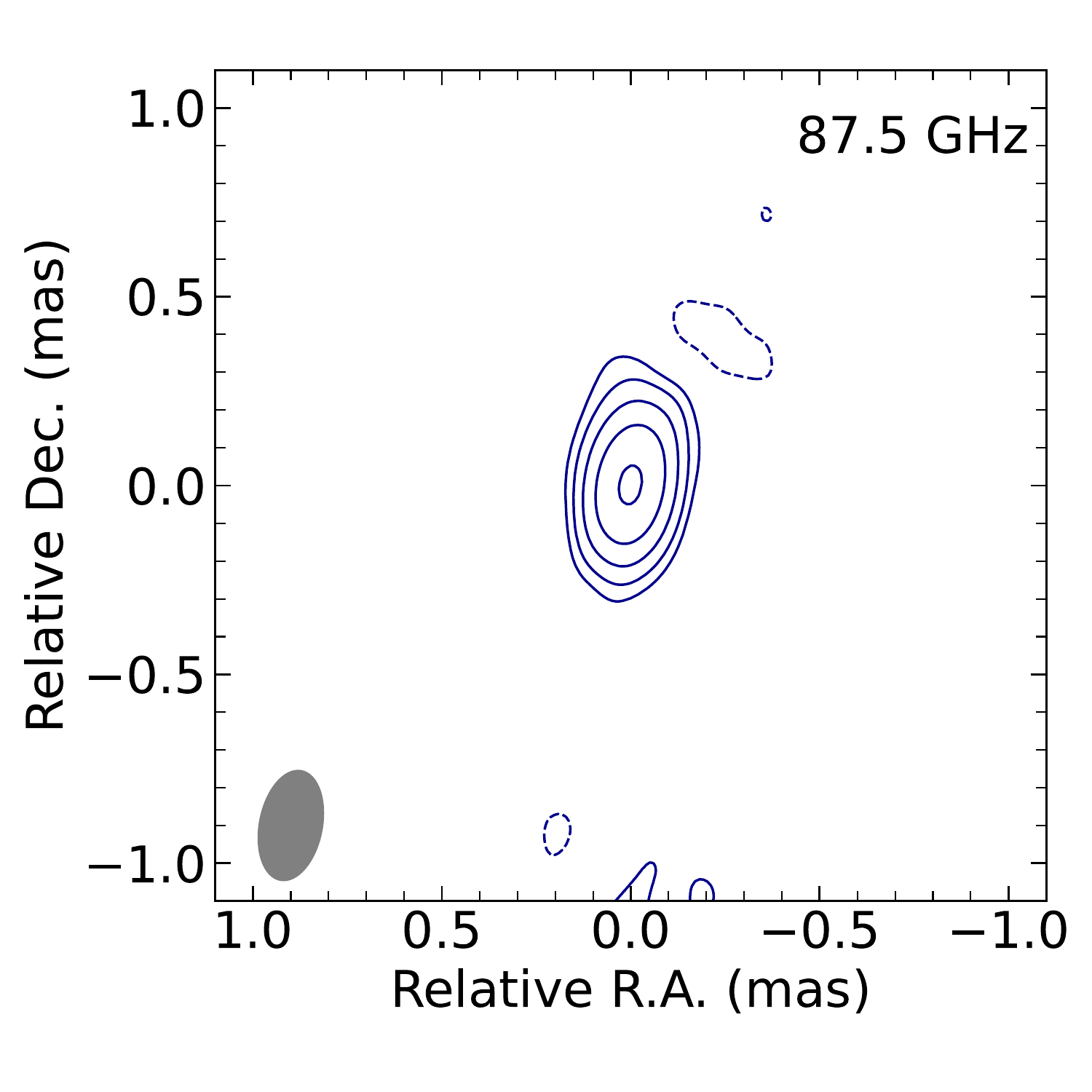}
\caption{Natural-weighted SFPR-ed VLBI image of M\,84 at 88\,GHz after self-calibration. The peak flux density is 38.1\,mJy\,beam$^{-1}$. The contour levels are at 2.2$\times(-1, 1, 2, 4, 8, 16)$\,mJy\,beam$^{-1}$. The lowest level is three times the rms noise of the image. The beam indicated in the left bottom corner is 0.29\,mas\,$\times$\,0.16\,mas at $-11^\circ$. \label{fig:fig1}}
\end{figure} 
\begin{figure}[htb]
\plottwo{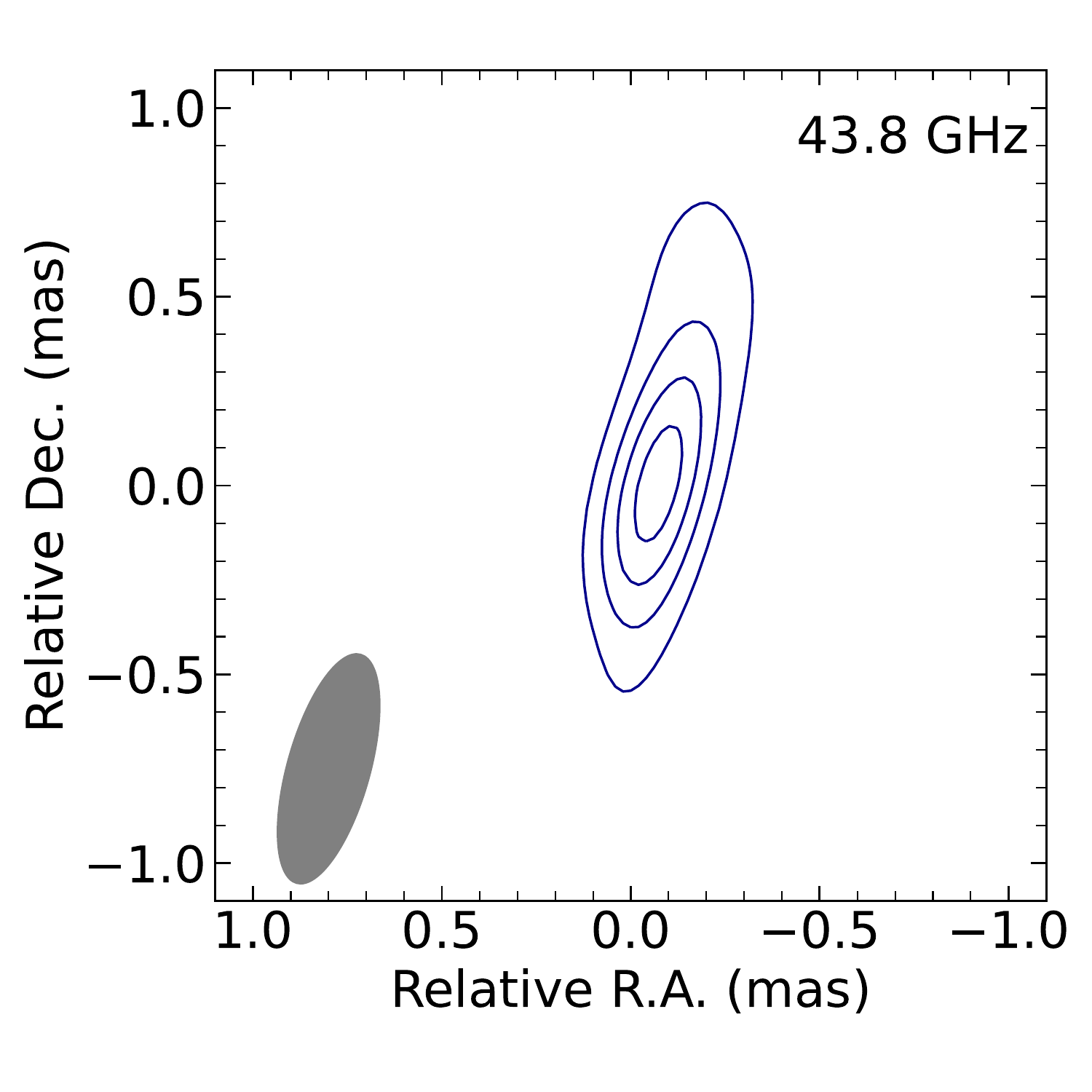}{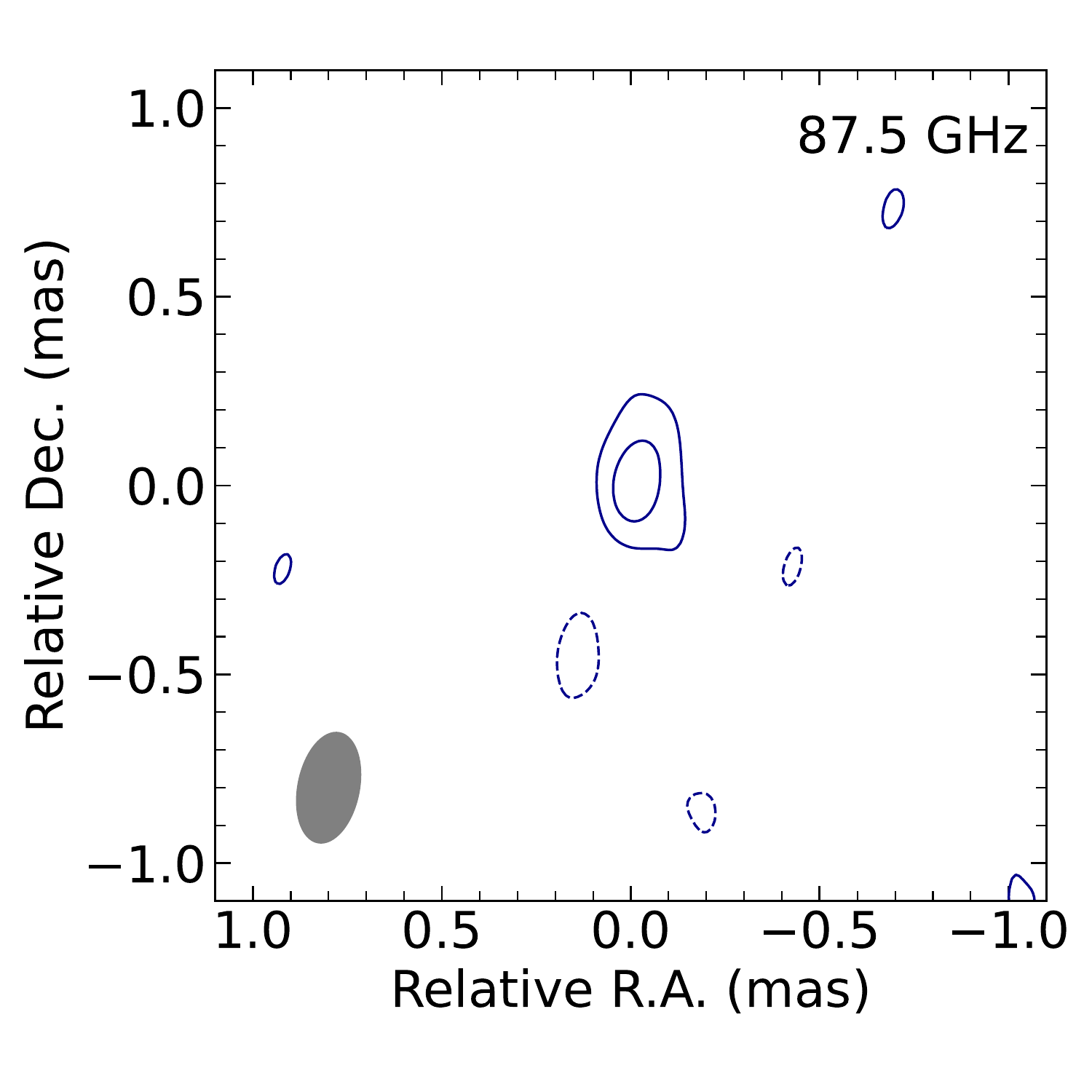}
\caption{SFPR-ed images of M\,84 at 44 (left) and 88\,GHz (right) on 2019 June 22 and 2021 March 31, respectively. The peak intensity is 4.1 and 8.2\,mJy\,beam$^{-1}$, respectively. The contour levels are at $-3$, 3, 6, 9 and 12 times the rms of the images (0.3 and 1.0\,mJy\,beam$^{-1}$). The full-width at half-maximum of the convolving beams (0.63\,mas\,$\times$\,0.22\,mas at $-15^\circ.6$ and 0.29\,mas\,$\times$\,0.16\,mas at $-11^\circ$.3) are shown in grey at the bottom-left corners of each image.\label{fig:fig2}}
\end{figure}
\begin{figure}[htb]
\plotone{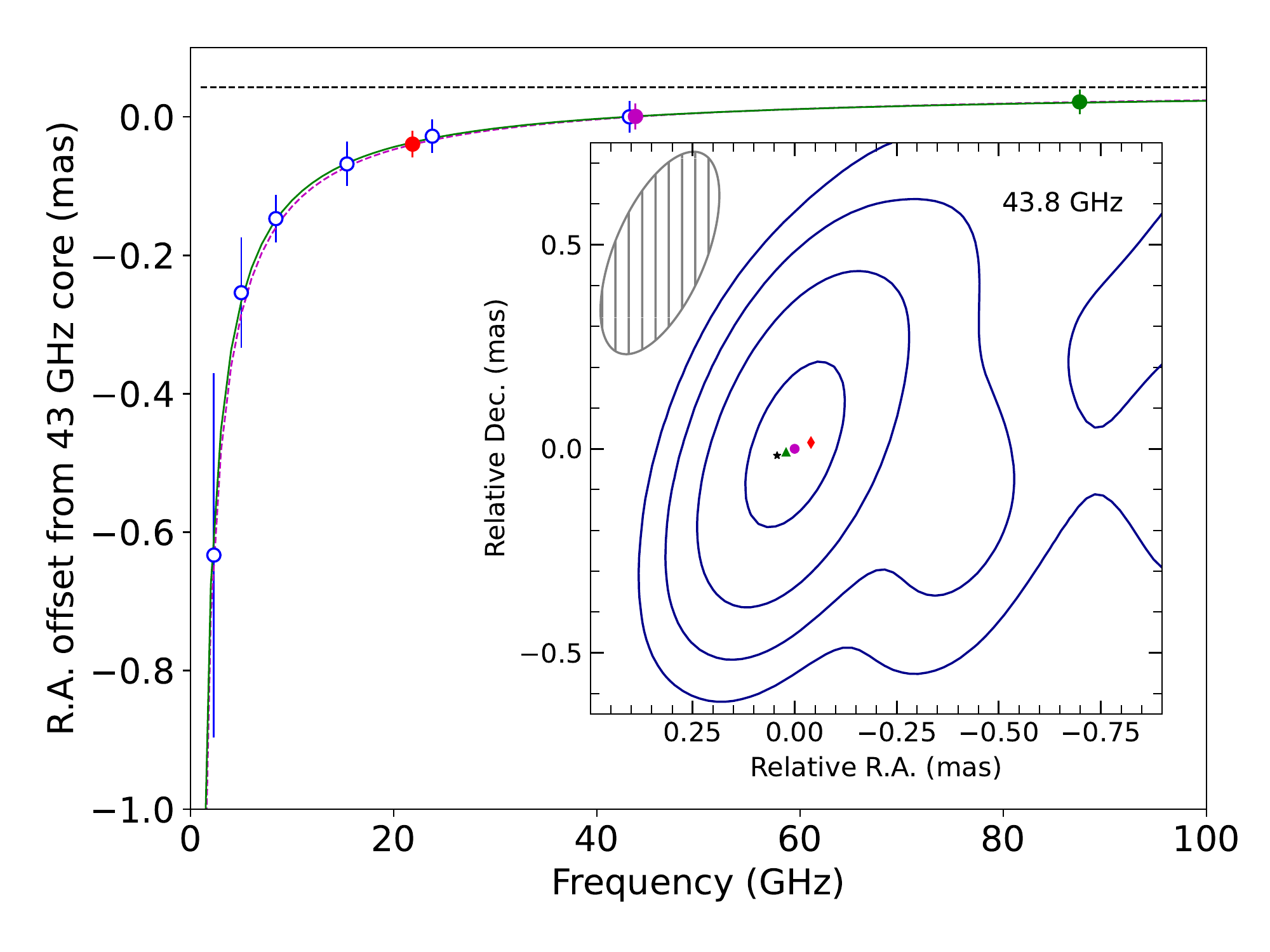}
\caption{Offsets in right ascension (RA) with respect to observing frequency from 43\,GHz core in M\,87. The data in hollow circles are from \citet{2011Natur...477..185H}. The solid circles are the measurements in this work. The fitting to the these two datasets with the same formula as \citet{2011Natur...477..185H} is the green curve, indicating  the jet apex position of $43$\,$\mu$as in RA (black dashed line). The fitting to the data of this work only is the dashed curve in magenta. The subplot in the bottom-right corner shows the core positions at 22, 44, 88\,GHz and the jet apex, which are marked as diamond, circle, triangle and star, respectively, overlapping with the VLBI image of M\,87 at 44\,GHz in 2021. The contour levels are at 12.0\,mJy\,beam$^{-1}$ ($27\sigma$ noise level of the image) stepping by a factor of 3, the beam in grey is 0.53\,mas\,$\times$\,0.22\,mas at $-22^\circ.7$. \label{fig:fig3}}
\end{figure}
\begin{deluxetable*}{ccccccc}
\tablenum{1}
\tablecaption{Summary of VLBI image parameters of M\,84 and M\,87 in the observations \label{tab:vlbi}}
\tablewidth{0pt}
\tablehead{
\colhead{} &\colhead{} &\colhead{} & \multicolumn{4}{c}{Synthesized beam on M\,84 / M\,87}
 \\ \cline{4-7}
\colhead{No.} & \colhead{Epoch} & \colhead{Frequency} &  \colhead{Major axis} & \colhead{Minor axis} & \colhead{Position angle} &  \colhead{RA component} \\
\colhead{} & \colhead{} & \colhead{GHz} &  \colhead{mas} & \colhead{mas} & \colhead{degree} &  \colhead{mas} 
}
\decimalcolnumbers
\startdata
1 &  2019-06-22 & 21.9 &  1.49 / 1.26 &  0.41 / 0.37 & $-14^\circ.5$ / $-20^\circ.5$&0.42 / 0.39    \\
2 & 2019-06-22 & 43.8 & 0.63 / 0.57 & 0.22 / 0.21 &  $-15^\circ.6$ / $-20^\circ.4$&0.23 / 0.22 \\
3 &  2021-03-31 & 43.8 &  0.60 / 0.53 &  0.24 / 0.22  &$-18^\circ.0$ / $-22^\circ.7$&0.25 / 0.24    \\
4 & 2021-03-31 & 87.5 & 0.29 / 0.27 & 0.16 / 0.15 & $-11^\circ.0$ / $-19^\circ.6$&0.16 / 0.16  \\
\enddata
\tablecomments{Col.(1) Number of VLBI images at different epochs and observing frequencies;  Cols.(2)-(3) Observing date and frequency; Cols.(4)-(6) The parameters (major axis, minor axis and position angle) of the nominal synthesized beam at each frequency; Col.(7) The beam size in the RA direction (RA component) of the nominal synthesized beam at each frequency.}
\end{deluxetable*} 
\begin{deluxetable*}{ccccc}
\tablenum{2}
\tablecaption{Fitting the location of jet apex in M\,87 \label{tab:vlbi}}
\tablewidth{0pt}
\tablehead{
\colhead{} &\colhead{} & \multicolumn{3}{c}{$r_{RA}=A\nu^{-k}+B$}
 \\ \cline{3-5}
\colhead{No.} & \colhead{Core-shift data used} &  \colhead{A} & \colhead{k} & \colhead{B} 
}
\decimalcolnumbers
\startdata
1 &  this work & $-1.45$ &  0.92 &  $0.045$    \\
2 & \citet{2011Natur...477..185H} and this work &  $-1.36\pm0.15$ & $0.92\pm0.06$ & $0.043\pm0.007$  \\
\enddata
\tablecomments{Col.(1) Number of the fitting;  Col.(2) Core-shift measurements used; Cols.(3)-(5) Fitting parameters $A$, $k$ and $B$. $B$ is the offset of the jet apex in RA direction from the 43\,GHz core.}
\end{deluxetable*} 


\end{document}